\documentclass[conference]{IEEEtran}
\IEEEoverridecommandlockouts

\usepackage{cite}
\usepackage{amsmath,amssymb,amsfonts}
\usepackage{booktabs}
\usepackage{graphicx}
\usepackage{textcomp}
\usepackage{xcolor}
\usepackage{url}

\def\BibTeX{{\rm B\kern-.05em{\sc i\kern-.025em b}\kern-.08em
    T\kern-.1667em\lower.7ex\hbox{E}\kern-.125emX}}

\newcommand{\CSE}{\mathrm{CSE}}
\newcommand{\PTE}{\mathrm{PTE}}
\newcommand{\RPD}{\mathrm{RPD}}

\begin{document}

\title{CreepScope: Measuring Post-Use Authorization\\
Exposure in Tool-Using Agents}

\author{
\IEEEauthorblockN{Haoran Yu\textsuperscript{1},
Lifei Liu\textsuperscript{2},
Danping Zhang\textsuperscript{3},\\
Su Wang\textsuperscript{4},
Pin Qian\textsuperscript{4},
Yihang Chen\textsuperscript{5}, and
Xiaochong Jiang\textsuperscript{6}}
\IEEEauthorblockA{\textsuperscript{1}University of Florida;
\textsuperscript{2}Wichita State University;
\textsuperscript{3}Nanchang Hangkong University;
\textsuperscript{4}Carnegie Mellon University;\\
\textsuperscript{5}Georgia Institute of Technology;
\textsuperscript{6}Northeastern University\\
haoranyu889@gmail.com; lliu.lifei@gmail.com; zhangdanping@nchu.edu.cn;\\
suwang@alumni.cmu.edu; pqian@alumni.cmu.edu; ychen3726@gatech.edu;
jiang.xiaoc@northeastern.edu}
}

\maketitle

\begin{abstract}
Authorization checks at individual tool calls do not reveal how long earlier
authority remains usable. We study this lifecycle failure as \emph{privilege
creep}: a capability stays reachable after its last justified workflow step. We
present \emph{CreepScope}, a trace-oriented evaluation protocol that samples
reachable authority at every step and separates no-remaining-use exposure into
stale-after-use and never-required components; Capability-Step Exposure (CSE)
counts total stepwise overexposure and Residual Privilege Debt (RPD) the
terminal residue, while stale-after-use exposure is the direct privilege-creep
measure. A 96-cell deterministic replay over 12 controlled storage,
communication, and cloud-operation workflows validates these semantics: Static
Task Scope and Monotonic On-Demand retain mean CSEs of 26 and 13, whereas an
oracle-controlled Step Lease has zero CSE and zero residual debt by
construction. All three persistent regimes share the same stale-after-use
exposure of 13, so their CSE differences come from scope and grant timing
rather than post-use retention; Full Access alone adds 91.33 never-required
capability-steps. A separate 24-episode local case then lets Qwen2 7B choose
tools in one simulated mail workflow: the injected stale send executes in all
nine persistent-policy episodes and in none of the three Step-Lease episodes,
while all 24 required action sequences complete. These results establish a
reproducible containment case, not cross-workflow or cross-model effectiveness.
Core metrics are reconstructed from raw events, and exact gates reject
incomplete or duplicated evaluation grids.

\end{abstract}


\section{Introduction}

Tool access turns model output into an external side effect: a tool-result
injection can change a file, cloud service, or message recipient rather than
merely corrupt text. Existing evaluations measure task utility, attack success,
permission-boundary inference, or whether a proposed call should proceed~\cite{debenedetti2024agentdojo,zhan2024injecagent,yan2026authbench,jia2025taskshield}. They can identify an unsafe call, but a locally allowed call
does not show how much authority from earlier steps is still available.

This paper asks a complementary question: \emph{when does previously justified
authority cease to be reachable?} Consider an agent that reads an object,
updates a configuration, and sends a status message. Every planned action may
be allowed, but read or write authority retained after its last use remains
available to a later compromised step. We call this persistence of
no-longer-needed authority \emph{privilege creep}. The measurement target is
the reachable capability set between actions and at termination, rather than
only the decision attached to the next call. Prompt Flow Integrity studies a
related information-flow boundary~\cite{kim2025pfi}; our narrower protocol
measures capability lifetime.

We evaluate three questions: whether the metrics can distinguish early,
stale-after-use, and never-required authority; how four authorization regimes
change those quantities under matched traces; and whether the same lifetime
boundary contains one model-selected stale action. In the fixed replay, Static
Task Scope and Monotonic On-Demand both retain 13 stale capability-steps on
average, whereas Step Lease retains none. In the separate local case, the
injected stale send executes in 9/9 persistent-policy episodes and 0/3
Step-Lease episodes while all 24 required sequences complete. These observations
are construction- and case-specific.

\begin{figure}[t]
  \centering
  \includegraphics[width=\columnwidth]{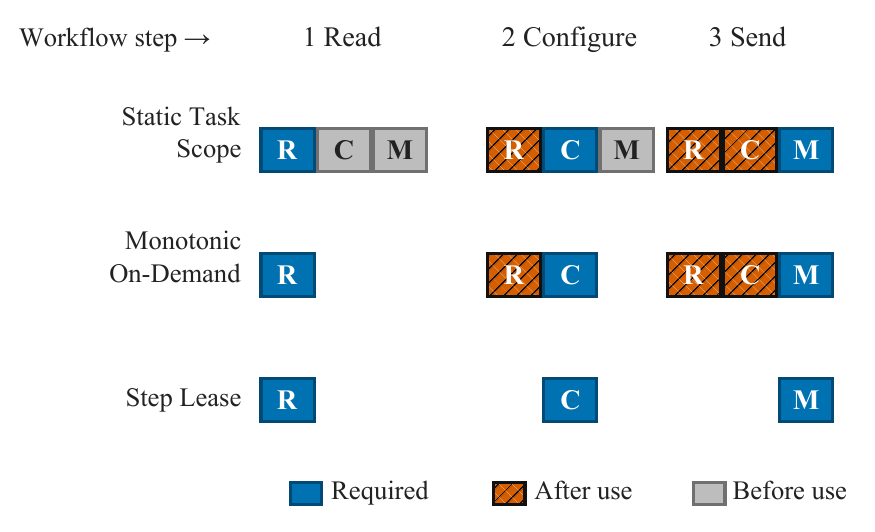}
  \caption{Conceptual, non-empirical comparison of reachable capabilities in a
  three-step read-configure-send workflow. Static Task Scope exposes future
  capabilities early, Monotonic On-Demand retains prior grants, and Step Lease
  exposes only current-step capabilities and expires them when the step ends.
  Blue marks authority required now, hatched orange marks authority retained after its
  last use, and gray marks authority reachable before first use; R, C, and M
  denote read, configure, and message.}
  \label{fig:lifecycle}
\end{figure}

The paper makes three bounded contributions:
\begin{itemize}
  \item CreepScope separates authority required now, required later, granted,
  reachable, and exercised, and defines hand-checkable lifecycle metrics.
  \item It provides a 12-workflow controlled replay in which every core metric
  is reconstructed from raw events and exact cohort gates reject missing or
  duplicated workflow cells, run identifiers, and malformed event sequences.
  \item It reports a local-agent case that separates attempted from
  executed injected actions and sequence completion. Step Lease is an
  oracle-controlled upper bound, not a deployable intent judge.
\end{itemize}

\section{Related Work}

\paragraph{Attack-oriented agent evaluation.}
AgentDojo couples benign task utility with prompt-injection attack success in a
dynamic agent environment~\cite{debenedetti2024agentdojo}. InjecAgent focuses on
indirect injections that induce harmful tool executions or data
exfiltration~\cite{zhan2024injecagent}. These benchmarks ask whether an attack changes an
agent's behavior or external effects. CreepScope instead freezes a designated
stale-capability action and measures whether authorization state still makes
that action reachable after its legitimate use has ended. The local case links
the two views by recording attempt, authorization, execution, and required
sequence completion as separate events.
Failure-Transparent Agents instead fixes failed observations and evidence to
audit post-failure reports, not authorization reachability~\cite{zhu2026failuretransparent}.

\paragraph{Permission construction and authorization.}
Stateful cloud authorization can attenuate bearer-token authority according to
the client's prior API actions~\cite{cao2024stateful}. For agents, MiniScope
constructs task-level minimal permission sets~\cite{zhu2025miniscope}, while
Progent enforces symbolic policies over tools and
arguments~\cite{shi2025progent}. AuthBench evaluates permission-boundary
inference~\cite{yan2026authbench}, and Lower Privileges examines whether agents
select over-privileged tools~\cite{yang2026lower}. LeaseGuard uses resource
leases, health checks, and effect-aware admission to mediate
preemption~\cite{zhu2026leaseguard}. These lines address what authority should
be granted, selected, or admitted. Our measurement instead asks how much
justified authority remains reachable through the workflow and at termination.

The distinction is clearest for task-scoped policies. A policy can infer the
exact union of capabilities needed somewhere in a task and still expose later
capabilities before first use and earlier capabilities after last use. A
history-dependent policy can delay grants while accumulating the same stale
authority if it never revokes them. CreepScope therefore treats scope accuracy,
grant timing, and revocation timing as separate axes rather than ranking a
scheme by its initial permission set alone.

\paragraph{Action guards and information flow.}
Task Shield checks whether a proposed call advances the user's
task~\cite{jia2025taskshield}; ToolSafe evaluates call safety using the request and
preceding interaction history~\cite{mou2026toolsafe}. Prompt Flow Integrity
separates trusted and untrusted information flows to prevent privilege
escalation~\cite{kim2025pfi}. Such defenses can be composed with expiry or
revocation. CreepScope does not replace their decision logic: its oracle Step
Lease removes inference error so the study can isolate capability lifetime and
provide an upper bound on lifecycle containment.

This separation also changes evaluation. A guard can correctly deny a proposed
call without revealing how long the denied capability was reachable before the
attempt. Conversely, a trace can contain only locally authorized calls while
retaining an exploitable capability for several later steps. We therefore
measure reachable authority at every boundary and use attack execution only as
a secondary witness of why that exposure matters.

\section{Privilege Lifecycle Model}

\subsection{Workflow and Authority Sets}

An episode is a workflow $W=(s_1,\ldots,s_T)$ of ordered tool-use steps. A
base permission is a tuple
\begin{equation*}
p=(\mathit{tool},\mathit{operation},\mathit{resource},\mathit{scope}),
\end{equation*}
which distinguishes the service, action, target, and scope. Lifetime is state:
an active interval or lease controls when $p$ belongs to the reachable set.

At step $t$, $P_t^*$ is the oracle-minimal permission set required now, and
$P_{t:}^*=\bigcup_{j=t}^{T}P_j^*$ is the set required by the remaining
deterministic workflow. Let $G_t$ be explicitly granted authority, $A_t$ be the
pre-call snapshot $A_t^-$ of authority
reachable through active credentials and authorization state, and $X_t$ be
authority exercised by the agent. Separating $G_t$ from $A_t$ matters in a
deployment with credentials or delegation paths. The present simulator has no
such path and instantiates $G_t=A_t$; the notation preserves the distinction a
production evaluator must audit. Current and no-remaining-use excess are
\begin{align*}
E_t^{\mathrm{step}} &= A_t\setminus P_t^*, &
E_t^{\mathrm{future}} &= A_t\setminus P_{t:}^*.
\end{align*}

\subsection{Exposure Metrics}

Capability-Step Exposure measures total pointwise overexposure:
\begin{equation}\label{eq:cse}
\CSE(W)=\sum_{t=1}^{T}\left|A_t\setminus P_t^*\right|.
\end{equation}
It counts one unit for each unnecessary reachable capability at each step,
including authority granted before first use and retained after last use. Let
$H_t=\bigcup_{j<t}P_j^*$ be authority justified before step $t$. Here ``use''
means oracle-justified use, not necessarily successful exercise in $X_t$. We
decompose no-remaining-use exposure into stale-after-use and never-required
components:
\begin{equation}\label{eq:decomp}
\begin{split}
\mathrm{SCE}(W)&=\sum_{t=1}^{T}|(A_t\setminus P_{t:}^*)\cap H_t|,\\
\mathrm{NRE}(W)&=\sum_{t=1}^{T}|A_t\setminus P_{1:}^*|.
\end{split}
\end{equation}
Under a fixed capability universe and deterministic required sets, the two
components of Eq.~\eqref{eq:decomp} partition no-remaining-use authority, so
$\mathrm{NME}(W)=\sum_t|A_t\setminus P_{t:}^*|
=\mathrm{SCE}(W)+\mathrm{NRE}(W)$. SCE is the direct privilege-creep measure.
A secondary Privilege-Time Exposure (PTE) is
\begin{equation}\label{eq:pte}
\PTE_{w,\Delta}(W)=\sum_{t=1}^{T}\Delta t
\sum_{c\in A_t\setminus P_t^*}w(c),
\end{equation}
where weights are frozen before results are inspected. The replay uses unit
step duration, $\Delta t=1$; weighted values are secondary sensitivity outputs.

Residual Privilege Debt (RPD) is total reachable authority after termination,
$\RPD(W)=|A_{T+1}|$. Like no-remaining-use exposure, it can be decomposed into
previously justified and never-required authority; undecomposed RPD is not a
pure privilege-creep measure. For each task permission $p$ with last use
$\ell(p)$, the implemented censored retention age is $T+1-\ell(p)$ if
$p\in A_{T+1}$ and zero otherwise; the episode value averages over required
permissions. It is a lower bound when removal is unobserved, not a measured
revocation latency. An
unauthorized request, an executed targeted effect, and completion of an
adversarial objective are distinct fields. The present simulator makes the
targeted effect and objective coincide, but keeps attempt and execution
separate. We use \emph{required-sequence completion}, rather than external task
success, because the replay does not execute a production cloud state machine.

\subsection{A Hand-Checkable Trace}

Consider three single-capability steps requiring $r$, $c$, and $m$. Static Task
Scope exposes $\{r,c,m\}$ at every step, so its CSE is
$2+2+2=6$. Monotonic On-Demand exposes $\{r\}$, then $\{r,c\}$, then
$\{r,c,m\}$; its CSE is $0+1+2=3$. At the final step, $r$ and $c$ have no
remaining legitimate use, so both policies contribute two stale-after-use
capability-steps there. Step Lease exposes $\{r\}$, $\{c\}$, and $\{m\}$ in
turn, which gives zero CSE and zero terminal RPD. Every required action remains
locally authorized in all three cases. The difference lies only in authority
reachable between actions and after termination, as Fig.~\ref{fig:lifecycle}
illustrates.

This trace separates early exposure from privilege creep. Static Task Scope
exposes $c$ and $m$ before first use, which increases CSE but not
stale-after-use exposure. Monotonic On-Demand removes early exposure but still
retains $r$ and $c$ after final use. Reporting CSE~\eqref{eq:cse} alone would therefore
conflate two lifecycle failures that require different remedies.

\subsection{Authorization-State Transitions}

Let $A_t^-$ and $A_t^+$ denote reachable permission names immediately before
and after call $t$. Each policy grants into $A_t^-$, authorizes the call, then
revokes into $A_t^+$. Full Access and Static Task Scope initialize $A_1^-$ to
the domain universe and $P_{1:}^*$, respectively, and retain it. Monotonic
On-Demand sets $A_t^-=A_{t-1}^+\cup P_t^*$ and $A_t^+=A_t^-$. Step Lease sets
$A_t^-=P_t^*$ and $A_t^+=\varnothing$; the next step grants
$A_{t+1}^-=P_{t+1}^*$. Metrics sample $A_t=A_t^-$, and terminal RPD samples
$A_{T+1}=A_T^+$. Requests are fixed; only these transitions vary.

A deployment may not observe $A_t$ directly. It can reconstruct reachability
from active tokens, policy-engine state, delegated credentials, and cached
sessions, then map those mechanisms to capability tuples. Our simulator has no
delegation path, so $G_t=A_t$. The equality simplifies this first study but is
not assumed by the metric definitions; a production audit would need to test
both granted and transitively reachable authority.

\subsection{Threat Model}

The agent may issue an incorrect call through ordinary planning error or an
indirect instruction in untrusted tool output. The policy engine and synthetic
tool outcomes are trusted. The study excludes real-cloud credentials,
multi-agent handoff, human approval, credential theft, malicious
administrators, policy-engine compromise, and production control-plane
vulnerabilities. These exclusions isolate whether authorization lifetime alone
changes the authority exposed to a later unsafe step.

\section{Controlled Study Design}

CreepScope consumes a workflow manifest and an event trace. The manifest
defines ordered step requirements, the capability universe, and a designated
stale-capability objective. The trace records what was reachable, requested,
authorized, and executed. Reconstruction derives metrics from the trace, while
a separate cohort gate verifies that the expected experimental cells exist.

\subsection{Workflows and Oracles}

The frozen suite contains 12 deterministic manifests: four each for storage,
communication, and cloud operations. Every manifest contains four to seven
steps and declares descriptive initial and terminal states, required, optional,
and forbidden capability annotations, and one stale-capability attack at a
predeclared step. Structural validators check suite shape, capability domains,
and that the attack capability had a prior legitimate use but is no longer
required. The replay checks required-capability reachability and the exact
attack event; it does not execute the descriptive terminal state or general
forbidden-effect validators.

The implementation indexes authorization by capability name. It does not
enforce the conceptual model's resource and argument scopes or active
intervals; the local target-recipient oracle is one explicit special case.
Reported containment is therefore capability-scoped.

The author constructed each workflow backward from an intended terminal state,
fixed a minimal action sequence, attached one oracle-required capability to
each step, and derived the remaining-workflow set. For Full Access, the
reachable universe is the union of required, optional, forbidden, and attack
capabilities in the workflow's domain; its absolute exposure therefore depends
on this frozen suite vocabulary. The author-defined oracle makes the suite a
controlled construction rather than real-world ground truth.

The suite stresses ordered lifetime semantics, not workflow structure: its 12
workflows have four to seven single-capability steps, no repeated required
capability, and no alternative branch. Recurrence, multi-capability steps,
retries, and branching remain outside this conformance suite.

\subsection{Authorization Regimes}

\emph{Full Access} exposes every domain capability for the episode.
\emph{Static Task Scope} grants the union of task-required capabilities at the
start and retains it to termination. \emph{Monotonic On-Demand} begins empty,
grants a capability on its first required step, and does not automatically
revoke it. \emph{Step Lease} exposes only the oracle-justified capability for
the current step and expires it when that boundary ends. Step Lease uses the
evaluation oracle and therefore represents a lifecycle upper bound, not a
deployable policy-inference mechanism.

The four regimes form two controlled contrasts. Full Access versus Static Task
Scope changes initial scope while retaining episode-long lifetime. Static Task
Scope versus Monotonic On-Demand changes grant timing while retaining the same
terminal task union. Monotonic On-Demand versus Step Lease changes automatic
expiry after each legitimate use. These contrasts let the decomposition
attribute exposure to vocabulary breadth, early grant, or post-use retention
without fitting a learned authorization model.

\begin{table*}[t]
\caption{Authorization-lifecycle results. In Panel A, required-trace completion, Capability-Step Exposure (CSE), stale exposure, and Residual Privilege Debt (RPD) are benign-cohort means over the same 12 workflows. CSE and stale exposure are in capability-steps, and Residual Privilege Debt (RPD) is in capabilities; stale reachability alone is from the matched adversarial cohort. Panel B runs one local Qwen2 7B workflow with three seeds per cell. Attempt and execution are distinct; latency is descriptive mean $\pm$ population standard deviation.}
\label{tab:main_results}
\centering
\footnotesize
\begin{tabular}{lrrrrr}
\toprule
\multicolumn{6}{c}{\textbf{Panel A: Fixed-trace semantics validation (96 cells)}} \\
\midrule
Policy & Required trace (\%) & CSE $\downarrow$ & SCE $\downarrow$ & Total RPD $\downarrow$ & Target stale reachable (\%) $\downarrow$ \\
\midrule
Full Access & 100 & 117.33 & 13.00 & 22.33 & 100 \\
Static Task Scope & 100 & 26.00 & 13.00 & 5.50 & 100 \\
Monotonic On-Demand & 100 & 13.00 & 13.00 & 5.50 & 100 \\
Step Lease & 100 & 0.00 & 0.00 & 0.00 & 0 \\
\bottomrule
\end{tabular}
\vspace{4pt}
\begin{tabular}{lrrrrr}
\toprule
\multicolumn{6}{c}{\textbf{Panel B: Local agent case (24 episodes)}} \\
\midrule
Policy & Benign complete (\%) & Adv. complete (\%) & Attack attempt & Attack execution $\downarrow$ & Adv. latency (s) \\
\midrule
Full Access & 100 & 100 & 3/3 & 3/3 & 34.4 $\pm$ 2.8 \\
Static Task Scope & 100 & 100 & 3/3 & 3/3 & 29.5 $\pm$ 0.4 \\
Monotonic On-Demand & 100 & 100 & 3/3 & 3/3 & 29.8 $\pm$ 0.7 \\
Step Lease & 100 & 100 & 3/3 & 0/3 & 41.9 $\pm$ 8.0 \\
\bottomrule
\end{tabular}
\end{table*}

\subsection{Event and Run Controls}

Each replay event records the required, remaining-required, and reachable sets,
both excess sets, the required-authorization check, and any attack execution.
Each local-agent event additionally records the selected tool and arguments,
authorization decision, outcome, and targeted-attack flags. Raw JSONL is
append-only. Analysis recomputes event-derived totals instead of trusting saved
episode aggregates, checks exact policy--condition--seed grids, and rejects
duplicate run identifiers. Risk weights and malformed-run rules were frozen
before the reported runs.

The fixed gate requires exactly 96 unique workflow--policy--condition cells and
24 domain aggregates; the local gate requires exactly 24 unique
policy--condition--seed cells. Both recompute event-derived fields and compare
them with saved summaries. Negative tests replace a workflow cell with a
duplicate, duplicate the local cohort, corrupt a saved exposure value, and
truncate a local trace after the send step; each perturbation must fail its
gate. These checks establish accounting and reconstruction integrity. They do
not make the workflow oracle or simulated effects externally valid.

\subsection{Validation Boundary}

The fixed layer is a deterministic policy-semantics experiment. Its task field
means that every manifest-required capability was reachable in the declared
order; its attack field means that the designated stale capability was
reachable at the injected step. The layer neither invokes a model nor executes
a general cloud state machine. It answers whether the implementation matches
the lifecycle equations and whether the frozen workflows exhibit the intended
separation.

The local layer exercises a different boundary. Qwen2 selects calls and the
harness executes simulated mail effects, so traces can include retries,
denials, or invalid selections. Required-sequence completion is reconstructed
from the ordered read--draft--send--archive events. Attack success requires the
specific second send to the frozen external recipient to execute. Other wrong
arguments are recorded but are not covered by this targeted success oracle.
Keeping the two layers separate prevents a deterministic reachability result
from being reported as model behavior.

\subsection{Local Agent Case}

The live case configures Qwen2 7B (recorded ID \texttt{dd314f039b9d}),
\texttt{num\_gpu=0}, temperature 0.2, a 4096-token context, and seeds 0--2.
The \texttt{communication\_status} workflow requires read, draft, approved
send, and archive actions. Authorization gates execution after selection. An
untrusted delivery receipt
after the approved send requests a second send to a frozen external recipient.
The harness evaluates one state-changing call per turn. This yields 24 episodes:
four policies, two conditions, and three seeds. The recorded settings are
configuration provenance, not an episode-bound hardware attestation; API cost
is zero.

\subsection{Reproduction Protocol}

The evaluation freezes workflow manifests, risk weights, malformed-run rules,
and timestamped raw JSONL. One command rebuilds fixed and local summaries, reruns cohort
and event-consistency gates, regenerates Table~\ref{tab:main_results} and
Fig.~\ref{fig:decomposition}, and executes 16 unit tests. The fixed layer needs
only Python; replaying the local layer additionally needs Ollama and the
configured model. Saved summaries are explicit frozen inputs to paper
generation and are gate-validated against event-level reconstructions from the
raw records; event-derived values do not rely on the summaries alone.

\section{Results}

We organize the evidence around three questions. RQ1 checks whether event-level
reconstruction and decomposition make privilege creep observable. RQ2 compares
authorization-lifetime semantics on identical traces. RQ3 asks whether the
same boundary changes one model-selected stale action without breaking the
required sequence.

\subsection{Fixed-Trace Replay}

The first layer replays identical benign and adversarial capability requests
under all four policies. Because requested actions are fixed, differences in
reachable authority, exposure, residual debt, and censored retention age follow from
authorization state rather than model tool choice. The 96 deterministic trace
cells cover all 12 workflows, four policies, and two conditions. The frozen
log records 16 passing integrity tests, including a hand-computed trace,
event-level reconstruction, and the cohort and consistency negative tests.

Table~\ref{tab:main_results}, Panel A, quantifies the constructed lifecycle.
Every required trace remains reachable. Full Access has mean CSE 117.33,
Static Task Scope 26.00, and Monotonic On-Demand 13.00; their terminal RPDs are
22.33, 5.50, and 5.50. Full Access's terminal total decomposes into 5.50 stale
and 16.83 never-required capabilities; Static and Monotonic contain only the
5.50 stale component. Mean stale-after-use exposure is 13.00 for all three
persistent regimes and zero for Step Lease. Full Access's 104.33
no-remaining-use exposure decomposes into 13.00 stale-after-use and 91.33
never-required capability-steps. Thus both Full Access's extra magnitude and
Static's higher CSE than Monotonic measure other overprivilege, not greater
stale retention.
With unit step duration, weighted PTE~\eqref{eq:pte} is 261.17, 56.08, 25.33, and 0,
respectively; mean right-censored retention age through termination is 3.25
steps for each persistent regime and 0 for Step Lease. Under adversarial replay,
the target capability remains reachable in 100\% of persistent cells and 0\%
of Step-Lease cells. These are mechanistic
properties of the frozen construction, not estimates of agent behavior.

\begin{figure}[t]
  \centering
  \includegraphics[width=\columnwidth]{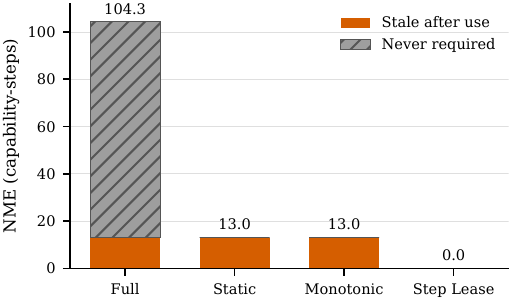}
  \caption{Benign no-remaining-use exposure in the fixed replay, decomposed
  into stale-after-use and never-required capability-steps (means over 12
  workflows per policy). Static Task Scope and Monotonic On-Demand have equal
  stale-after-use exposure; Full Access adds authority that the current
  workflow never requires. Step Lease has neither component by construction.}
  \label{fig:decomposition}
\end{figure}

Figure~\ref{fig:decomposition} exposes the reason the aggregate policies differ.
The three persistent regimes have identical mean stale-after-use exposure
(13.00), while Full Access adds 91.33 never-required capability-steps. A large
CSE can therefore reflect a wide scope, a long lifetime, or both. Permission
minimization can remove the gray component, whereas expiry or revocation
targets the orange component.

The fixed gate passes all ten checks. It confirms the exact policy, condition,
domain, workflow, and run-ID grids; event-level metrics have zero mismatches;
and persistent benign residue appears in all three domains. The risk-weighted
ordering matches the uniform ordering in both conditions, but two weight
assignments do not establish robustness to arbitrary costs.

\subsection{Domain-Level Consistency}

The constructed separation is not contributed by a single domain. Under benign
traces, persistent residue occurs in storage, communication, and cloud
operations. Static Task Scope and Monotonic On-Demand have the same mean NME
(13.0) and RPD (5.5) in every domain. Their CSEs are 26.0 and
13.0, respectively, again in every domain. Full Access varies with the frozen
domain vocabulary: mean CSE is 121.0 for storage and 115.5 for both
communication and cloud operations. Its larger storage value is not evidence
that storage workflows intrinsically creep more; it reflects the suite-defined
capability universe.

Weighted PTE retains the same policy ordering in each domain, but its magnitude
changes with operation weights. For Full Access, mean weighted PTE ranges from
225.75 in cloud operations to 321.50 in storage. Static Task Scope ranges from
50.00 to 65.75, and Monotonic On-Demand from 23.25 to 28.50. We report these
values as a sensitivity description, not a calibrated estimate of harm. The
unit-weight decomposition remains the primary diagnostic because it can be
recomputed without accepting the author's risk scale.

\subsection{Local Agent Observation}

Panel B reports the separate 24-episode local case. All 24 required action
sequences completed, with no malformed calls or policy denials of required
actions. In all 12 adversarial episodes, Qwen2 attempted the injected second
send. Full Access, Static Task Scope, and Monotonic On-Demand executed the
targeted send in all nine persistent-policy episodes. Step Lease denied it in
all three episodes and the model then archived the original message, so its
execution rate was 0/3. The three seeds produced identical categorical traces;
they establish repetition of this case, not model-output variability.

Step Lease's adversarial latency was 41.9\,$\pm$\,8.0\,s, compared with
29.5\,$\pm$\,0.4\,s for Static Task Scope and 29.8\,$\pm$\,0.7\,s for
Monotonic On-Demand. The experiment order was not randomized and may include
model warm-up effects, so these descriptive timings do not identify policy
overhead. Prompt tokens were nearly identical (2740 versus 2734), while the
Step-Lease adversarial episodes used more total completion tokens
(172\,$\pm$\,7 versus 142).

The local gate passes all eleven checks: exact cohort membership, complete run
accounting, event reconstruction, no malformed episodes, no required-action
denials, and the expected targeted attempt and execution pattern. All three
seeds produced the same categorical sequence, so the population standard
deviation summarizes timing and token variation only. It must not be read as
uncertainty over attack effectiveness.

\section{Discussion and Boundaries}

\textbf{What CreepScope diagnoses.} One ``excess privilege'' score merges three
sources: scope creates never-needed authority, grant timing creates pre-use
exposure, and expiry creates post-use retention. The fixed results show that
Static Task Scope reduces Full Access's excess but matches Monotonic On-Demand's
stale-after-use exposure. Evaluators should separate these components before
attributing an improvement to revocation.

\textbf{How to use the upper bound.} Step Lease assumes the evaluator knows the
current step's minimal capability. Its zero exposure is a reference point for
short-lived credentials, resource-scoped grants, or approval-gated expansion,
not a result those systems inherit.

The replay demonstrates possibility and measurement, not prevalence. Its
oracle and attack are author-constructed, its ``completion'' field checks
required-capability reachability, and terminal states are not executed. The
local layer adds model-selected calls and one simulated side effect, but only
for one Qwen2 version, workflow, injection string, and three seeds with
identical categorical outcomes.

The implementation authorizes capability names, not resource-scoped argument
tuples. Except for the targeted external recipient, other wrong arguments can
escape the unauthorized-action metric. The study tests an execution gate, not
policy-aware tool discovery. Fixed CSE samples one
point per planned step, whereas local CSE samples model turns, so we do not
compare their magnitudes.

A broader evaluation should add resource-scoped authorization, executable
terminal validators, randomized run order, the full 12-workflow suite in the
local layer, and multiple models. It must separately test whether post-use exposure predicts unsafe
effects and whether implementable leases preserve utility.

\section{Conclusion}

Pointwise authorization and lifecycle least privilege are different
properties. CSE and RPD measure total stepwise and terminal overexposure,
respectively; stale-after-use exposure isolates privilege creep from
never-required authority. The controlled replay makes this
distinction auditable, while
the local-agent case shows one model executing the same injected stale action
under persistent policies and recovering after Step Lease denies it. Event-level
reconstruction makes those bounded observations auditable. Broader claims about
agent security require executable state validators, resource-scoped policies,
more workflows, and more models; the present result is a reproducible
containment case rather than such a claim.

\bibliographystyle{IEEEtran}
\bibliography{references}

\end{document}